\documentstyle[aps,epsfig,twocolumn]{revtex}

\draft

\begin{document}
\twocolumn[\hsize\textwidth\columnwidth\hsize\csname
@twocolumnfalse\endcsname
\title{\bf Relaxation dynamics of a linear molecule in a random static medium:\\
A scaling analysis}

\author{Angel J. Moreno $^{1}$\footnote{e-mail: angel.moreno@phys.uniroma1.it} \hspace{0.05 cm}
and  Walter Kob $^{2}$\footnote{Corresponding author; e-mail: kob@ldv.univ-montp2.fr}}

\address{$^{1}$ Laboratoire des Verres. Universit\'{e} Montpellier II, Place E. Bataillon,
CC 069, F-34095 Montpellier, France.}

\address{$^{2}$ Dipartimento di Fisica and INFM
Udr and Center for Statistical Mechanics and Complexity, Universit\`{a}
di Roma ``La Sapienza'', Piazzale Aldo Moro 2, I-00185 Roma, Italy.} 
\maketitle

\begin{abstract}
We present extensive molecular dynamics simulations of the motion of a
single linear rigid molecule in a two-dimensional random array of fixed
overlapping disk-like obstacles. The diffusion constant for the center of mass translation,
$D_{\rm CM}$, and for rotation, $D_{\rm R}$, are calculated for a
wide range of the molecular length, $L$, and the density of obstacles,
$\rho$.  The obtained results follow a master curve $D\rho^{\mu} \sim
(L^{2}\rho)^{-\nu}$ with an exponent $\mu = -3/4$ and 1/4 for $D_{\rm R}$
and $D_{\rm CM}$ respectively, that can be deduced from simple scaling
and kinematic arguments.  The non-trivial positive exponent $\nu$
shows an abrupt crossover at $L^{2}\rho = \zeta_{1}$. For $D_{\rm CM}$
we find a second crossover at $L^{2}\rho = \zeta_{2}$.  The values of
$\zeta_{1}$ and $\zeta_{2}$ correspond to the average minor and major
axis of the elliptic holes that characterize the random configuration
of the obstacles. A violation of the Stokes-Einstein-Debye relation is
observed for $L^{2}\rho > \zeta_{1}$, in analogy with the phenomenon of
enhanced translational diffusion observed in supercooled liquids close
to the glass transition temperature.

\end{abstract}
] 
\narrowtext

\section{Introduction}

Since it was initially introduced by Lorentz as a model for
the electrical conductivity in metals \cite{lorentz}, the
problem of the Lorentz gas has given rise to a substantial
theoretical effort aimed to understand its properties
\cite{bruin,alder,gotzea,gotzeb,gotzec,masters,keyes,machta,binder}.
In this model, a {\it single} classical particle moves through a
disordered array of {\it static} obstacles. It can thus be used as
a simplified picture of the motion of a light atom in a disordered
environment of heavy particles having a much slower dynamics.
It has been shown that, for high densities of obstacles, this model
presents many of the features of the dynamics of supercooled liquids
or dense colloidal systems, such as a transition from an ergodic phase
of non-zero diffusivity to a non-ergodic phase with diffusivity zero
\cite{bruin,alder,gotzea,gotzeb,gotzec,keyes}.  Despite of the absence
of a dynamics of the host medium, theoretical approaches to the Lorentz
gas problem are highly non-trivial. In particular, diffusion constants
and correlation functions are found to be non-analytical functions of
the density of obstacles \cite{bruin,gotzea,gotzeb,gotzec,machta,binder}.

Diffusing particles and obstacles are generally modeled as disks, spheres
or hyperspheres in two, three, and $n > 3$ dimensions, respectively.
In this paper we present an investigation by means of molecular dynamics
simulations of the relaxation dynamics of a generalization of the
Lorentz gas, in that the diffusing particle is given by a rigid rod.
This model can thus be seen as a simplified picture of the dynamics
of a linear molecule in a porous medium or in a colloidal suspension
of heavy particles. Furthermore this system will allow to gain insight
into the relaxation dynamics of a nonspherical probe molecule immersed
in a simple liquid, an experimental technique that is used, e.g.,
in photo-bleaching experiments to study the dynamics of glass-forming
liquids~\cite{cicerone95a,cicerone95b}, or the investigation of the
coupling between the rotational and translational degrees of freedom in
such liquids~\cite{andreozzi97,taschin01}.

For simplicity, the present simulations have been done in two dimensions, though
we expect that physical arguments similar to those presented here can
be used for the analysis of the results in the more general three-
or $n$-dimensional case.

The paper is organized as follows: The details of the investigated model
and the simulation are given in Section II.  Qualitative information about
the trajectories of the rod is presented in Section III.  Section IV presents
results for the rotational and center-of-mass translational diffusion
constants of the rod. An interpretation of the obtained results in terms
of characteristic lengths of the host medium is given in Section V and
the conclusions are presented in Section VI.

\section{Model and details of the simulation}

The generalization of the Lorentz gas we consider is a rigid rod
diffusing in a two-dimensional array of static disk-like obstacles.
The rod, of total mass $M$, consists of a string of $N$ linearly
aligned beads of equal mass $m=M/N$ with a bond length $2\sigma$.
The rod length is therefore given by $L=(2N-1)\sigma$.  Although from
the point of view of a theoretical calculation the use of hard-core
like interactions is preferable, it is simpler to do a simulation
with a continuous potential. Thus, we modeled the interaction between
single beads and single obstacles by a truncated and shifted soft-disk
potential, $V(r) = \epsilon {[}(\sigma/r)^{12}-(\sigma/r_{\rm c})^{12}{]}$
for $r<r_{\rm c}$, and zero for  $r>r_{\rm c}$.  As cutoff radius
we took a value of $r_{\rm c}=2.5\sigma$.  The obstacles were placed
randomly in a square box of length $l_{\rm box}$, which was used for
periodic boundary conditions. 
Obstacles were allowed to overlap, i.e., no minimal distance
was imposed between the centers of neighbouring disks.
The density of obstacles is defined as
$\rho = n_{\rm obs}/l_{\rm box}^{2}$, with $n_{\rm obs}$ the number
of obstacles. 
Subsequently a rod was randomly placed in the box and
was equilibrated at a temperature $T = \epsilon/k_{\rm B}$ for a time
longer than the relaxation time for the corresponding $\rho$ and $L$.
(Note that this equilibration was necessary since interactions are not
hard-core.)  In the following, space and time will be measured in the
reduced units $\sigma$ and $(\sigma^{2}m/\epsilon)^{1/2}$, respectively.

Once the rod was equilibrated, a production run was performed in the
microcanonical ensemble. The equations of motion of the rod were
formulated in terms of the coordinates of its center of mass and its
orientation. Due to the linear geometry of the rod, the constant length
of the bonds and equal mass of the beads, the total force and the torque
of the constraining forces on each single bead is zero, and thus only the
total force and the torque due to the external forces of the obstacles
on the beads are needed in the equations of motion. The latter
were integrated by using the velocity form of the Verlet algorithm. We
used a time step between 0.01 and 0.002, respectively for the smallest
and the largest investigated $L$ and $\rho$.  Production runs covered
typically about $10^6$ time units.  For each configuration of obstacles
we simulated 100 independent rods and in order to take the average over
the frozen disorder we investigated typically 6-10 different realizations
of obstacles. Thus, for each density and rod length we simulated between
600-1000 independent rods. The number of obstacles was typically 50000.

\section{Trajectories}

Fig.~\ref{fig1} shows for different rod length typical trajectories
of the center of mass of the rod over a time of $10^6$ time units and
the density $\rho=10^{-2}$. For short rods, $L=3$, the trajectory looks
like the one expected for a diffusive Brownian particle.  With increasing
$L$ the nature of the motion changes not only quantitatively but also
qualitatively.  As it is recognized from the figure for the case $L=79$,
for long rods the trajectory is composed of vertices in which the rod
rattles back and forth and long segments in which moves in a quasi-linear
way.  The reason for this change in the dynamics is that, for large $L$,
the rods can be temporarily trapped by the neighboring obstacles, an
effect that is well-known in the case of the original Lorentz gas for
$L=1$ and large $\rho$. For this latter system an increase of $\rho$
leads to a strong increase of the trapping time and hence to a slowing
down of the dynamics~\cite{bruin,alder}, in analogy to the mechanism that
leads to the dramatic slowing down of the relaxation dynamics in simple
glass-forming liquids upon cooling. This trapping mechanism is of course
also present in the case of the rods. However, while for the original
Lorentz gas, i.e., for disks, the diffusion dynamics is isotropic also
on small length scales, the translational relaxation dynamics of rods
in the direction perpendicular to their axis is strongly hindered by
the steric hindrance induced by the neighboring obstacles, leading
to the observed quasi-linear paths, in which the velocity of the rod
is approximately parallel to its axis and thus allowing for a quick
propagation. Although these events are relatively rare, they allow the
rod to move rapidly through the obstacles and hence they are important
for its relaxation dynamics. Since the time it takes for the rod to have
a noticeable velocity component that is (essentially) parallel to its
orientation and the length of the subsequent quasi-ballistic flight
is a random variable (due to the strong disorder of the obstacles),
the resulting motion is very erratic and also strongly heterogeneous
(i.e.  it depends strongly on the initial position and orientation of
the rod). Hence it is no surprise that the resulting average dynamics
is also very heterogeneous which can be seen, e.g., by the presence of
strongly non-Gaussian effects in the relaxation dynamics at large length
scales ~\cite{moreno}.

Fig.~\ref{fig1} also shows that the typical spatial extension of the
trajectory within a given time depends strongly on $L$ which implies that
also the translational diffusion constant for the center of mass shows a
strong $L-$dependence, in agreement with the results that we will discuss
below. Hence it might be suspected that the change in the nature of the
trajectories upon a change in $L$, see Fig.~\ref{fig1}, is just due
to the decrease in $D_{\rm CM}$. That this is, however, not the case,
can be seen in Fig.~\ref{fig2}, where we show two typical trajectories
for two rods that have a very different length but, by having made an
appropriate choice of the corresponding densities $\rho$, a very similar
translational diffusion constant for the center of mass. As can be seen,
the nature of the trajectories does indeed depend strongly on $\rho$
and $L$, even if the diffusion constant is the same.

\section{Diffusion constants and scaling laws}

We now investigate the $L-$ and $\rho-$dependence of the diffusion
constants for rotation, $D_{\rm R}$, and for translation of the
center of mass, $D_{\rm CM}$.  These quantities can be calculated  as
$D_{\rm R}=\langle(\Delta\phi)^{2}\rangle/2t_{\rm sim}$ and $D_{\rm
CM}=\langle(\Delta r)^{2}\rangle/4t_{\rm sim}$, respectively, where
$\langle(\Delta\phi)^{2}\rangle$ and $\langle(\Delta r)^{2}\rangle$ are
the mean square angular and center of mass translational displacements
at $t = t_{\rm sim}$, the time at the end of the simulation.

We first consider $D_{\rm R}$ and we assume that its dependence on
$L$ and $\rho$ is given by a product of the form 

\begin{equation}
D_{\rm R} = k_{\rm R}L^{-a}\rho^{-b} \quad , 
\label{eq0}
\end{equation}

\noindent
with $k_{\rm R}$ a constant prefactor. By using scaling and kinematic
arguments we show in the following that this Ansatz does indeed gives a
good description of the data in that we can generate a master curve for
$D_{\rm R}(L,\rho)$. To this aim we note that in the limit of infinitely
thin rods and point obstacles, i.e., in the limit that finite size effects
are absent, the geometry of the system is invariant if it is scaled by an
(arbitrary) factor $f$. Thus, since the length of the rod and
the density of obstacles for the scaled system will be given respectively
by $fL$ and $f^{-2}\rho$, it is clear that the relevant quantity for
the geometry of the dynamics is the dimensionless variable $L^2\rho$,
since it is invariant under scaling. It is therefore useful to rewrite the
Ansatz for $D_{\rm R}$ in Eq.~(\ref{eq0}) as:

\begin{equation}
D_{\rm R} = k_{\rm R}\rho^{\alpha}/(L^{2}\rho)^{\beta} \quad .
\label{eq1}
\end{equation}

\noindent
The corresponding rotational diffusion constant for a given system,
$D_{\rm R}$, and that of the rescaled system, $D_{\rm R}^{(f)}$
will therefore be related by

\begin{equation}
D_{\rm R}^{(f)} = f^{-2\alpha}D_{\rm R} \quad .
\label{eq2}
\end{equation}

In the case that $\alpha \neq 0$, the presence of the factor
$f^{-2\alpha}$ implies that, although the two systems are equivalent
from a {\it geometrical} point of view, this is not the case for their
{\it dynamics}, i.e., for the time scales for their relaxation. In order
to obtain the value of the exponent $\alpha$ we proceed as follows: A
characteristic time scale can be defined by $\tau = \psi/\omega$, where
$\psi$ is a typical angular displacement between consecutive collisions
and $\omega$ is the angular velocity in the absence of obstacles. Note
that, due to its geometric nature, $\psi$ does not depend on $L^{2}\rho$
if the finite extension of the obstacles and the width of the rod can
be neglected. In contrast to this, the angular velocity is not invariant
under scaling, as we show now.  If we approximate the discrete string
of beads by a continuous rod whose mass is distributed uniformly with
density $\lambda=M/L$, one obtains for the moment of inertia of the
rod $I=\lambda L^{3}/12$. From the equipartition theorem for the energy
we then find that $\langle \omega^{2}\rangle = 12k_{\rm B}T/\lambda
L^{3}$, i.e., $\omega \propto L^{-3/2}$.  Therefore, the time scales
for the original system and that of the rescaled system are related
as $\tau^{(f)} = f^{3/2}\tau$, and thus we find for the corresponding
rotational diffusion constants $D_{\rm R}^{(f)} = f^{-3/2}D_{\rm R}$.
From Eq.~(\ref{eq2}) we obtain therefore $\alpha=3/4$, and
Eq.~(\ref{eq1}) is hence transformed into the scaling law:

\begin{equation}
D_{\rm R}\rho^{-3/4} = k_{\rm R}(L^{2}\rho)^{-\beta} \quad ,
\label{eq3}
\end{equation}

\noindent
where the dynamic exponent $\beta$ and the prefactor $k_{\rm
R}$ are the only non-trivial quantities. The reliability of the
scaling law~(\ref{eq3}) is controlled by two factors. One the one
hand, it depends on the accuracy of the factorization Ansatz of
Eq.~(\ref{eq1}).  Higher order logarithmic or power corrections
can be present when deducing diffusion constants from reptation ideas
\cite{doia,doib,fixman,teraoka,szamel} or, from first-principle theories,
as has been shown for the translational diffusion constant in the original
Lorentz gas in the framework of kinetic or Mode-Coupling-Theory (MCT)
\cite{bruin,gotzea}.  In the probable case where such corrections are
present, the exponent $\beta$ must be seen as an {\it effective} exponent
introduced as a fit parameter to describe a more complicated functional
form of $D_{\rm R}(L,\rho)$.  On the other hand, it also depends on the
accuracy of the approximation that the diameter of the obstacles and the
width of the rod can be neglected.  One certainly must expect that this
approximation is no longer valid once some characteristic length scale
of the host medium, such as the typical distance between neighboring
obstacles, becomes comparable to the thickness of the rod and obstacles.

Fig.~\ref{fig3} shows the calculated values of $D_{\rm R}\rho^{-3/4}$
for the different investigated lengths and densities as a function of
$L^2\rho$. It can be seen that, within a factor 300 in density, and
for all the rod lengths whose relaxation times can be accessed in the
time window of the simulation, the scaling law (\ref{eq3}) is nicely
fulfilled.  Furthermore, we see that the obtained master curve shows a
marked crossover at $L^{2}\rho = \zeta_{1} \approx 1.4$, from an exponent
$\beta=1.8$ to $\beta=2.7$, thus indicating a change in the rotational
relaxation dynamics when the length of the rod becomes larger than
$(\zeta_{1}/\rho)^{1/2}$. In the next section we will give a physical
interpretation of this dynamic crossover.

Similar arguments can be used to obtain the $L-$and $\rho-$dependence
of the translational diffusion constant of the center of mass. For this
we introduce an Ansatz that is analogous to Eq.~(\ref{eq1}):

\begin{equation}
D_{\rm CM} = k_{\rm CM}\rho^{\gamma}/(L^{2}\rho)^{\delta} \quad .
\label{eq4}
\end{equation}

\noindent
The same arguments that led to Eq.~(\ref{eq2}) can now be used to obtain the
relation between the translational diffusion constant of a given system
and the one for a system which has been obtained by rescaling the former
by a factor $f$:

\begin{equation}
D_{\rm CM}^{(f)} = f^{-2\gamma}D_{\rm CM}  \quad .
\label{eq5}
\end{equation}

\noindent
The corresponding characteristic time scale is defined as $\tau = l/v$,
with $l$ a typical distance between consecutive collisions and $v$ the
velocity of the center of mass of the rod in the absence of obstacles.
From the equipartition theorem for the energy we find $\langle v^{2}\rangle = 2k_{\rm
B}T/\lambda L$, i.e., $v \sim L^{-1/2}$.  Therefore, the characteristic
time will scale as $\tau^{(f)} = f^{3/2}\tau$, leading to  $D_{\rm
CM}^{(f)} = f^{1/2}D_{\rm CM}$.  Thus, from Eq.~(\ref{eq5}) we
have $\gamma=-1/4$, and Eq.~(\ref{eq4}) is transformed into the
scaling law:

\begin{equation}
D_{\rm CM}\rho^{1/4} = k_{\rm CM}(L^{2}\rho)^{-\delta} \quad .
\label{eq6}
\end{equation}

Fig.~\ref{fig4} shows the calculated values of $D_{\rm CM}\rho^{1/4}$
for the investigated densities and rod lengths (left set of data).
As for $D_{\rm R}$, a crossover is observed at $L^{2}\rho = \zeta_{1} =
1.4$, indicating a change also in the translational dynamics for rods
longer than $(\zeta_{1}/\rho)^{1/2}$.  Though a systematic shift is
obtained for increasing $\rho$, a description in terms of the scaling law
(\ref{eq6}) is reasonably good for $\rho < 10^{-2}$, where values of $D_{\rm
CM}\rho^{1/4}$ change by less than a factor of 2 for a variation of
density by more than a factor of 100. If one considers only the range
$L^{2}\rho < \zeta_{1}$, also the data for $\rho=3\cdot 10^{-2}$ follows
the master curve quite nicely. In any case, it seems that the validity
of the approximations used to obtain Eq.~(\ref{eq6}) -factorization
Ansatz and neglecting thickness of rods and obstacles- is more limited
for the translational dynamics than for rotations.

We also mention that a better collapse of the data is obtained if
we replace $\rho^{1/4}$ by $\rho^{0.35}$, see right set of data in
Fig.~\ref{fig4}.  However, we do not really give an important physical
meaning to this latter value of the exponent since it could just be the
result of taking into account a correction to the trivial value 1/4 by
lumping together finite size effects or higher order contributions to
the right side of Eq.~(\ref{eq6}).

Except for the highest investigated density $\rho = 3\cdot10^{-2}$, the
data in the range of $L^{2}\rho$ from $\zeta_{1} = 1.4$ to $\zeta_{2} =
22 $ also seem to follow the scaling law~(\ref{eq6}), as can be seen in
the inset in Fig.~\ref{fig4}, where the dynamic exponent $\delta=0.4$
is obtained from the best fit that fulfills continuity with the law
$\delta=0.7$ obtained for $L^{2}\rho < \zeta_{1}$.  Finally, as shown
by the breakdown of the scaling, it is clear that, at least for $\rho >
6\cdot10^{-3}$, the factorization Ansatz and/or neglecting finite size
effects are no more good approximations for $L^{2}\rho > \zeta_{2}=22$,
suggesting a further change in the translational dynamics for rods longer
than $(\zeta_{2}/\rho)^{1/2}$. In the next section we will come back to
this observation.

Some insight about the coupling between the rotational and translational
degrees of freedom can be obtained from investigating the validity
of the Stokes-Einstein-Debye relation $D_{\rm CM}\tau_{R}=c$, where
$c$ is a constant and $\tau_{R}$ the rotational relaxation time
\cite{egelstaff}.  We define this latter quantity as the time where
the correlation function $\langle\cos\Delta\phi(t)\rangle$  decays to
$e^{-1}$.  Here $\Delta\phi(t)$ is the angular displacement at time $t$
from the initial orientation of the rod at $t=0$, and brackets denote
the ensemble average. First we deduce a scaling law for $\tau_{\rm R}$
by introducing a factorization Ansatz $\tau_{\rm R} = k_{\tau_{\rm
R}}\rho^{\sigma}(L^{2}\rho)^{\eta}$. Thus, it is found that $\tau_{\rm
R}^{(f)}=f^{-2\sigma}\tau_{\rm R}$, where $\tau_{\rm R}^{(f)}$ is the
rotational relaxation time in the system that has been scaled by a
factor of $f$. As argued above, we also have the relation $\tau_{\rm
R}^{(f)} = f^{3/2}\tau_{\rm R}$. Therefore $\sigma=-3/4$ and we obtain
the scaling law

\begin{equation}
\tau_{\rm R}\rho^{3/4} = k_{\tau_{\rm R}}(L^{2}\rho)^{\eta} \quad .
\label{eq7}
\end{equation}

\noindent
As can be seen in Fig.~\ref{fig5}, data fulfill the latter
equation, with a crossover from an exponent $\eta=0.7$ to $\eta=2.7$
at $L^{2}\rho=\zeta_{1}=1.4$, i.e., at the same value of $L^{2}\rho$
as the crossover observed for the rotational diffusion constant, again
indicating a change in the rotational dynamics for rods longer than
$(\zeta_{1}/\rho)^{1/2}$.

From Eqs.~(\ref{eq6}) and (\ref{eq7}) we now obtain a scaling law
for the product $D_{\rm CM}\tau_{\rm R}$:

\begin{equation}
D_{\rm CM}\tau_{R}\rho = k_{\rm SED}(L^{2}\rho)^{\eta-\delta} \quad .
\label{eq8}
\end{equation}

\noindent
Thus, if for a given density $\rho$ the Stokes-Einstein-Debye relation is
fulfilled, the dynamic exponents $\delta$ and $\eta$, for the scaling laws
of respectively $D_{\rm CM}$ and $\tau_{\rm R}$, must cancel each other.
This cancellation is indeed fulfilled within the error bar for $L^{2}\rho <
\zeta_{1}$ as shown in Fig.~\ref{fig6}. (Recall that we have obtained 
$\delta=0.7$ from Fig.~\ref{fig4}, and $\eta=0.7$ from Fig.~\ref{fig5}.) Thus,
the Stokes-Einstein-Debye relation is violated for rods longer than
$(\zeta_{1}/\rho)^{1/2}$, in analogy with the phenomenon of enhanced
translational diffusion observed in supercooled liquids -with temperature
as control parameter-, close to the glass transition \cite{fujara,ediger}.

\section{Discussion}

\begin{figure}
\centering
\includegraphics[width=0.45\textwidth]{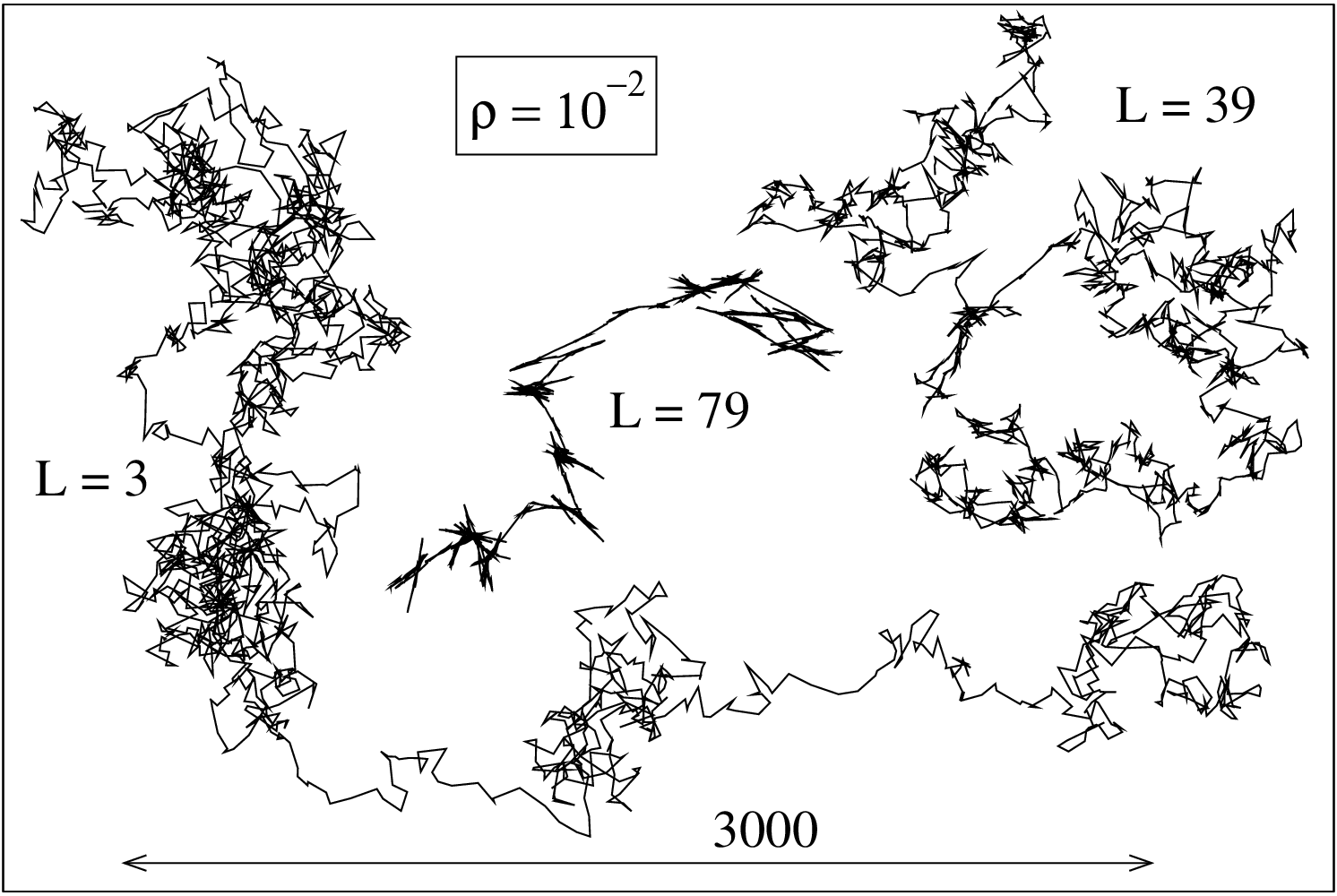}
\vspace{0.3 cm}
\caption{\label{fig1} Typical trajectories of the center of mass of a rod
for $L=$ 3, 39 and 79 at the density $\rho = 10^{-2}$. All them correspond
to a simulation time $t_{\rm sim} = 10^{6}$, with a time interval of 500
between consecutive plotted points. The arrow at the bottom indicates the 
length scale.}
\end{figure}

\begin{figure}
\centering
\includegraphics[width=0.45\textwidth]{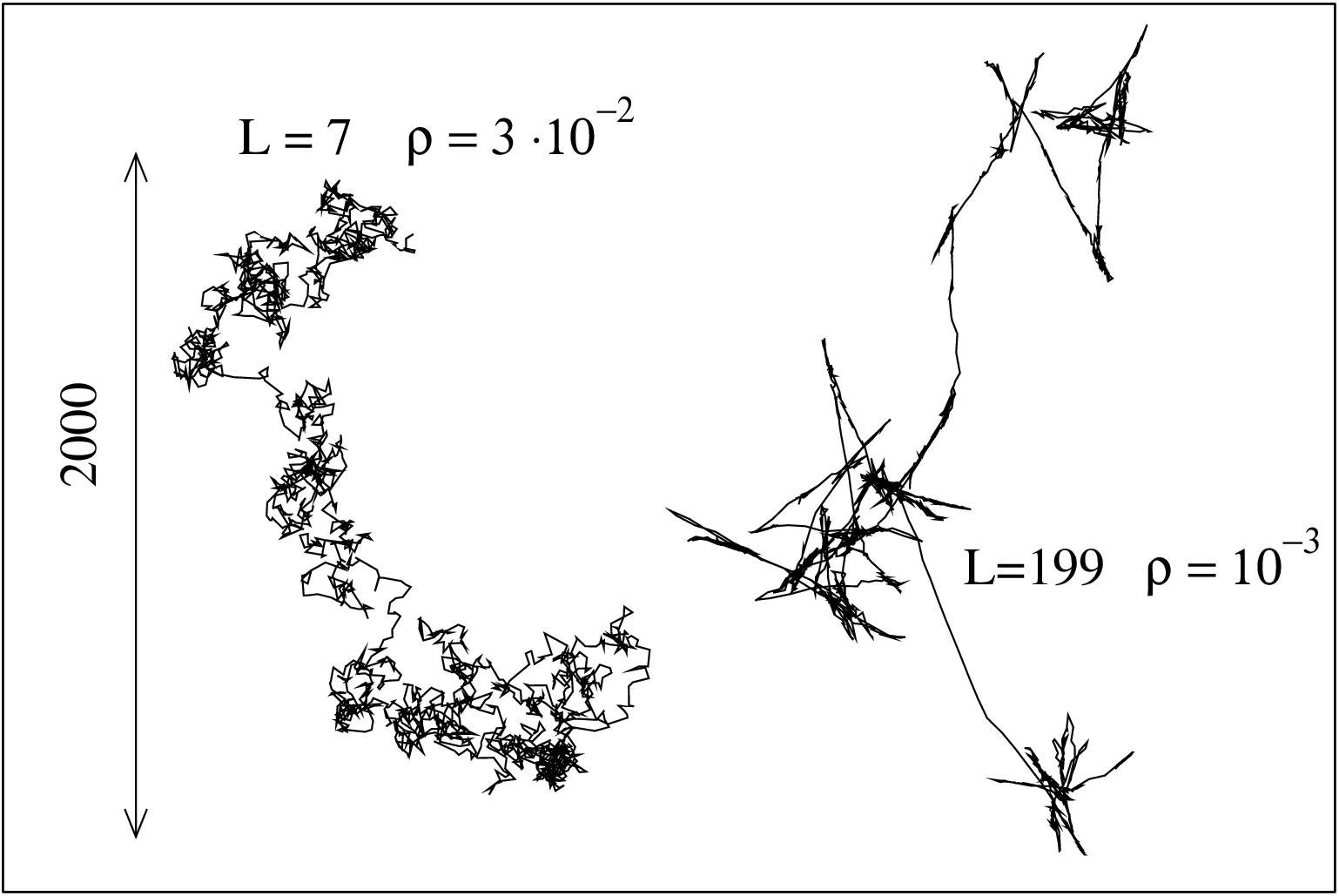}
\vspace{0.5 cm}
\caption{\label{fig2} Typical trajectories of the center of mass of the rod
for $L=7$, $\rho = 3 \cdot 10^{-2}$ and for $L=199$, $\rho = 10^{-3}$.
The simulation time and time interval between consecutive plotted points
are the same as in Fig.~\ref{fig1}.}
\end{figure}

\begin{figure}
\centering
\includegraphics[width=0.45\textwidth]{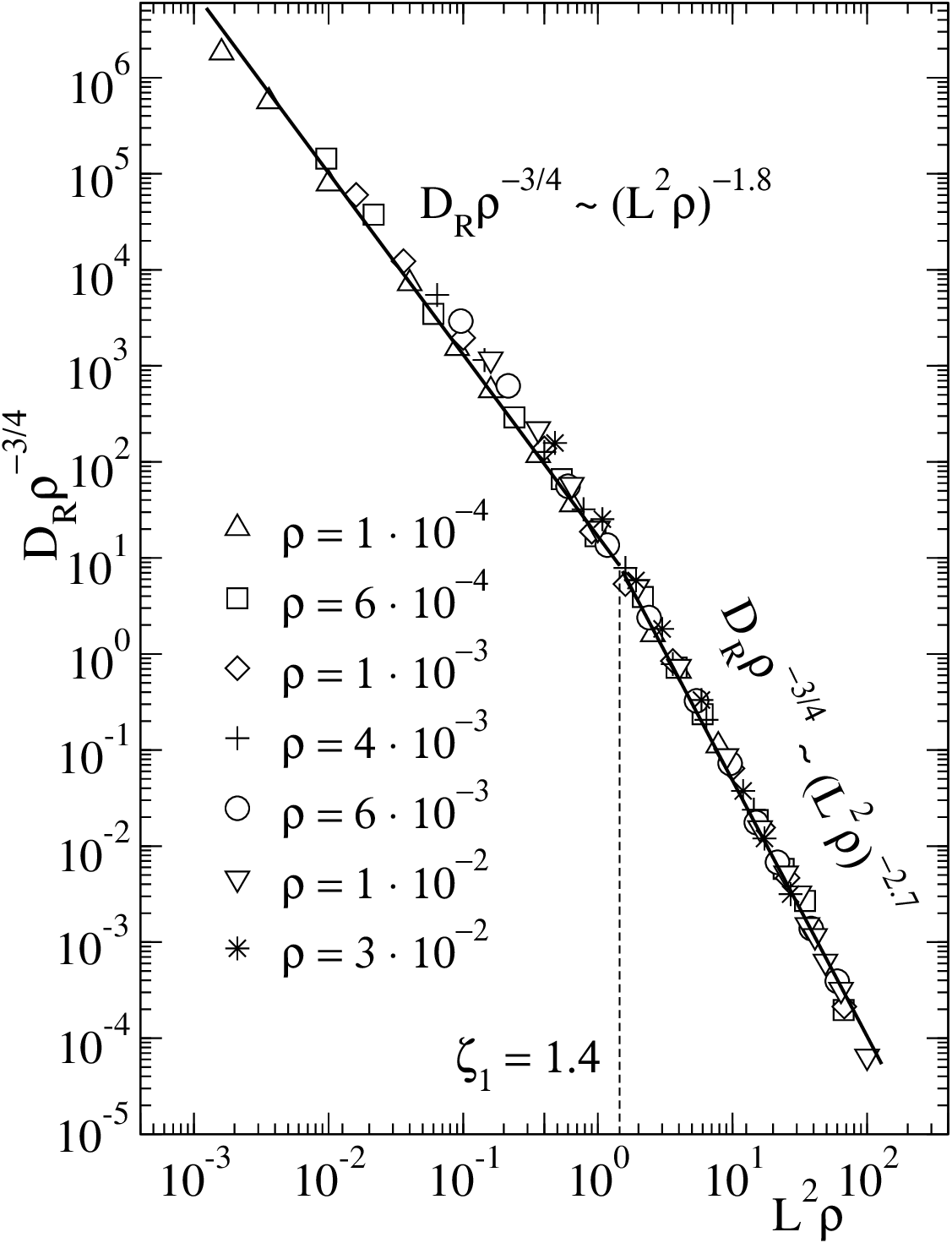}
\vspace{0.5 cm}
\caption{\label{fig3} Scaling law for the rotational diffusion
constant. The solid lines are fits to power laws.
The dynamic crossover at $L^{2}\rho = \zeta_{1}$ is indicated.}
\end{figure}

\begin{figure}
\centering
\includegraphics[width=0.45\textwidth]{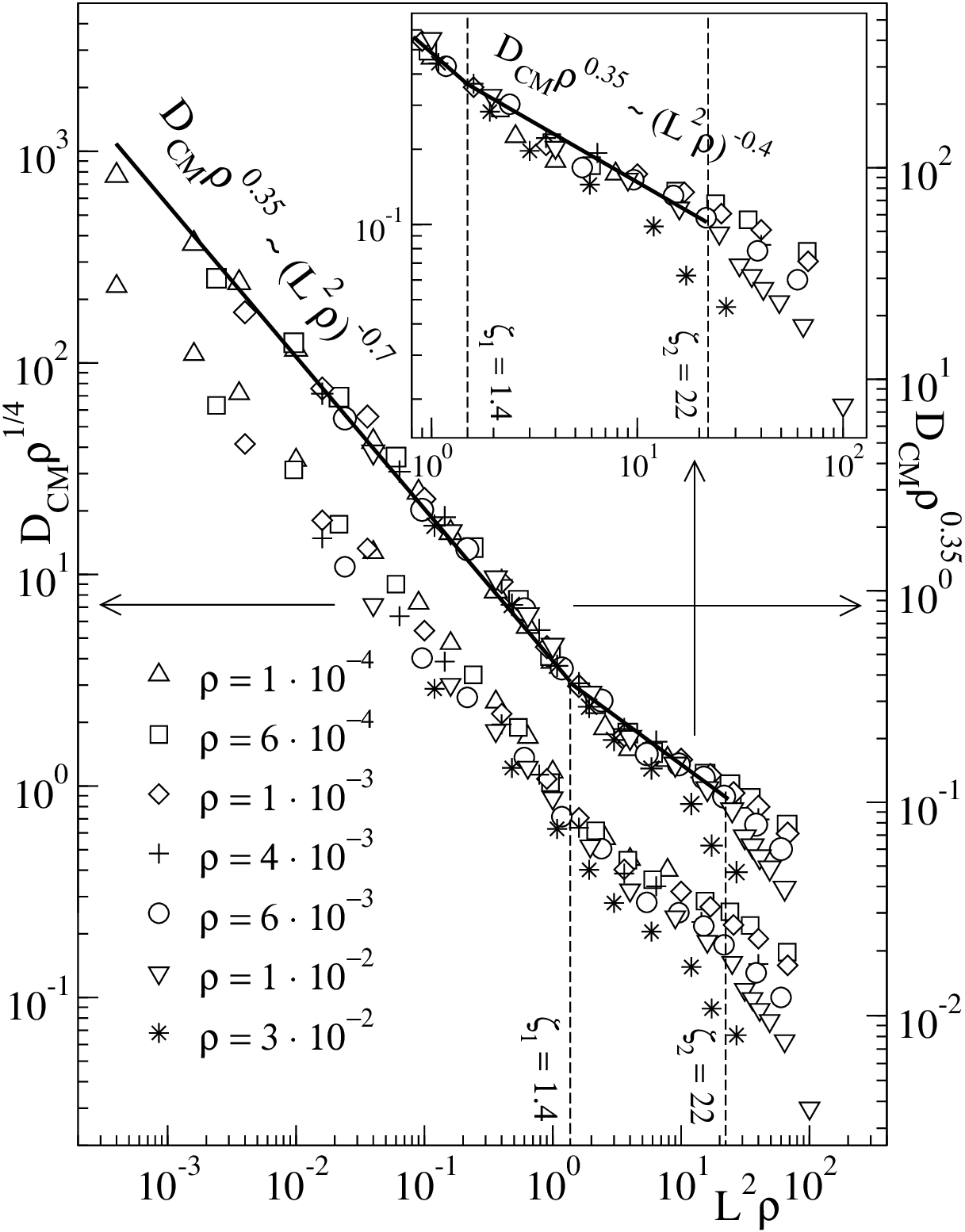}
\caption{\label{fig4} As Fig.~\ref{fig3} for the translational
diffusion constant of the center of mass. A double $y$-axis representation
is used for clarity. The inset shows a magnification of the data
$D_{\rm CM}\rho^{0.35}$.}
\end{figure}

\begin{figure}
\centering
\includegraphics[width=0.45\textwidth]{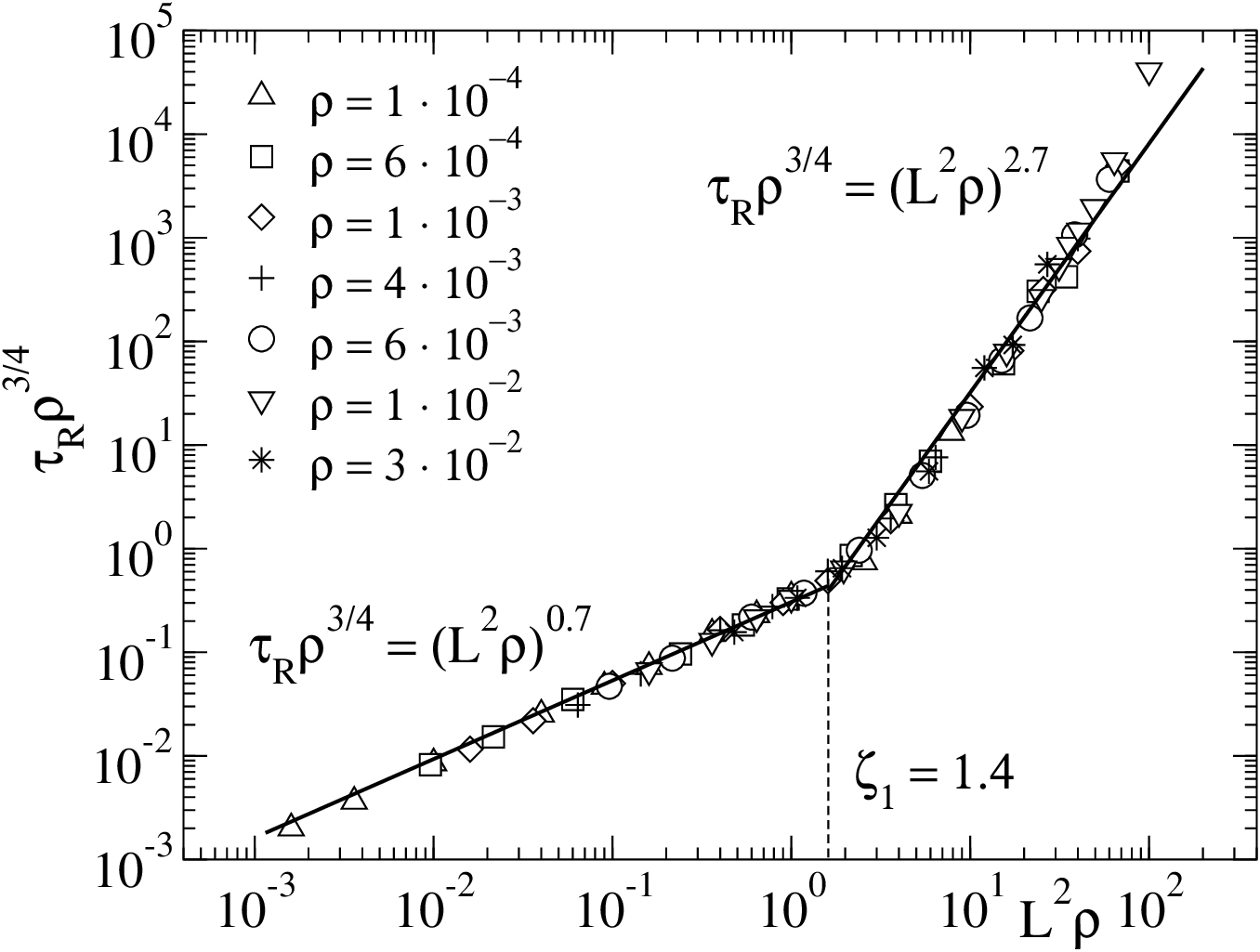}
\vspace{0.6 cm}
\caption{\label{fig5} Scaling law for the rotational relaxation time.
Solid lines are fits to power laws.}
\end{figure}

\begin{figure}
\centering
\includegraphics[width=0.45\textwidth]{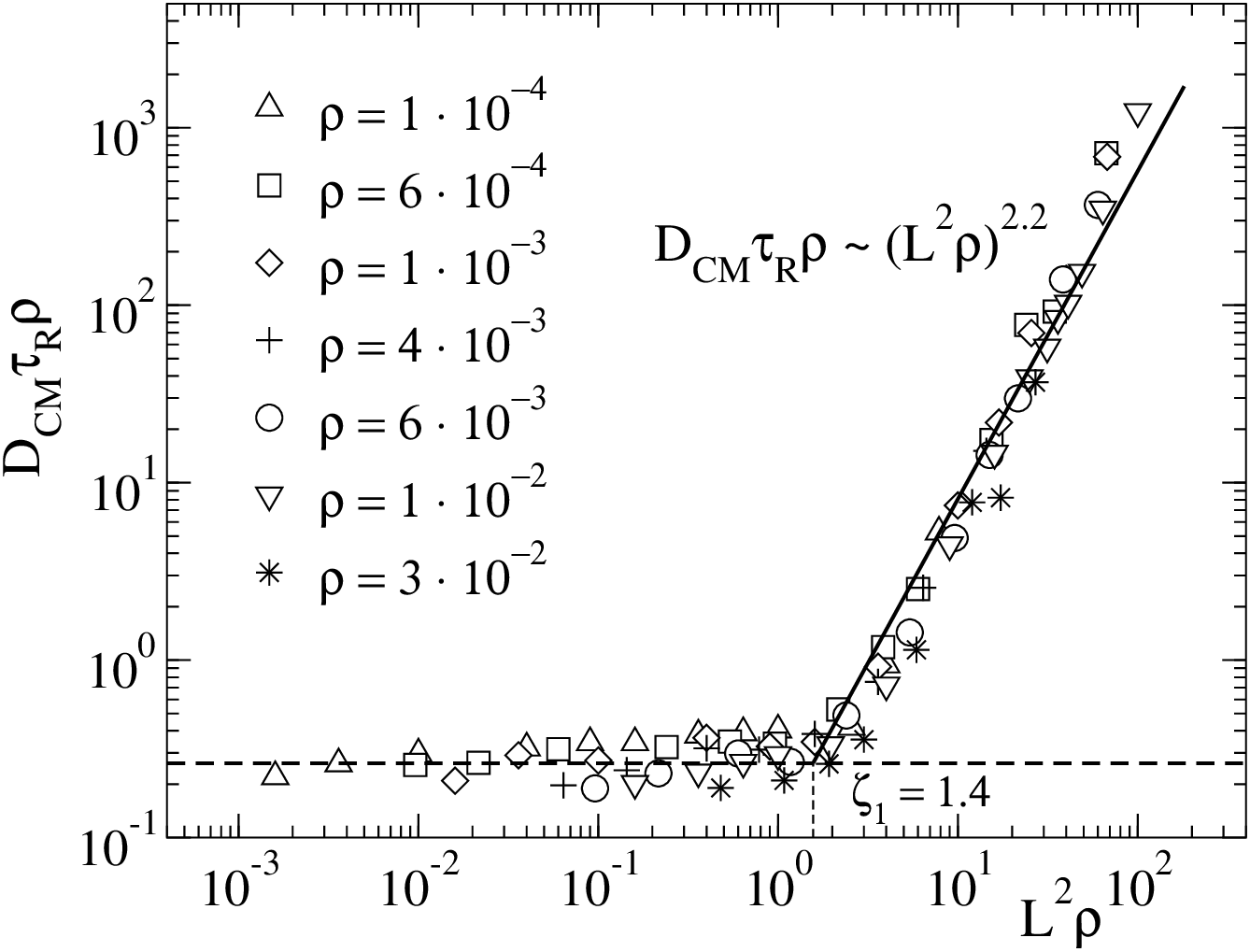}
\vspace{0.6 cm}
\caption{\label{fig6} Scaling law for the product $D_{\rm CM}\tau_{\rm R}\rho$
and violation of the Stokes-Einstein-Debye relation at large $L^{2}\rho$.
The dashed horizontal line marks a constant value. The solid line is a
fit to a power law.}
\end{figure}

From the investigation of the rotational diffusion constant and relaxation
time, we have observed two different dynamic regimes for rods shorter and
longer than a crossover length $\ell_{1}=(\zeta_{1}/\rho)^{1/2}$, with
$\zeta_{1}=1.4$. i.e., $\ell_{1}=1.2\rho^{-1/2}$.  This crossover has also
been observed for the translational dynamics. Moreover, for the latter,
a second crossover, apparently not present for the rotational dynamics
(see Fig.~\ref{fig3}), has been obtained.  The corresponding crossover
length is $\ell_{2}=(\zeta_{2}/\rho)^{1/2}$, with $\zeta_{2}=22$, i.e.,
$\ell_{2}=4.7\rho^{-1/2}$.  For the density $\rho=10^{-2}$, corresponding
to the trajectories represented in Fig.~\ref{fig1}, the crossover lengths
are $\ell_{1}=12$ and $\ell_{2}=47$.  Therefore, the latter trajectories
correspond to rods in the three dynamic regimes separated by $\ell_{1}$
and $\ell_{2}$.

\begin{figure}
\centering
\includegraphics[width=0.39\textwidth]{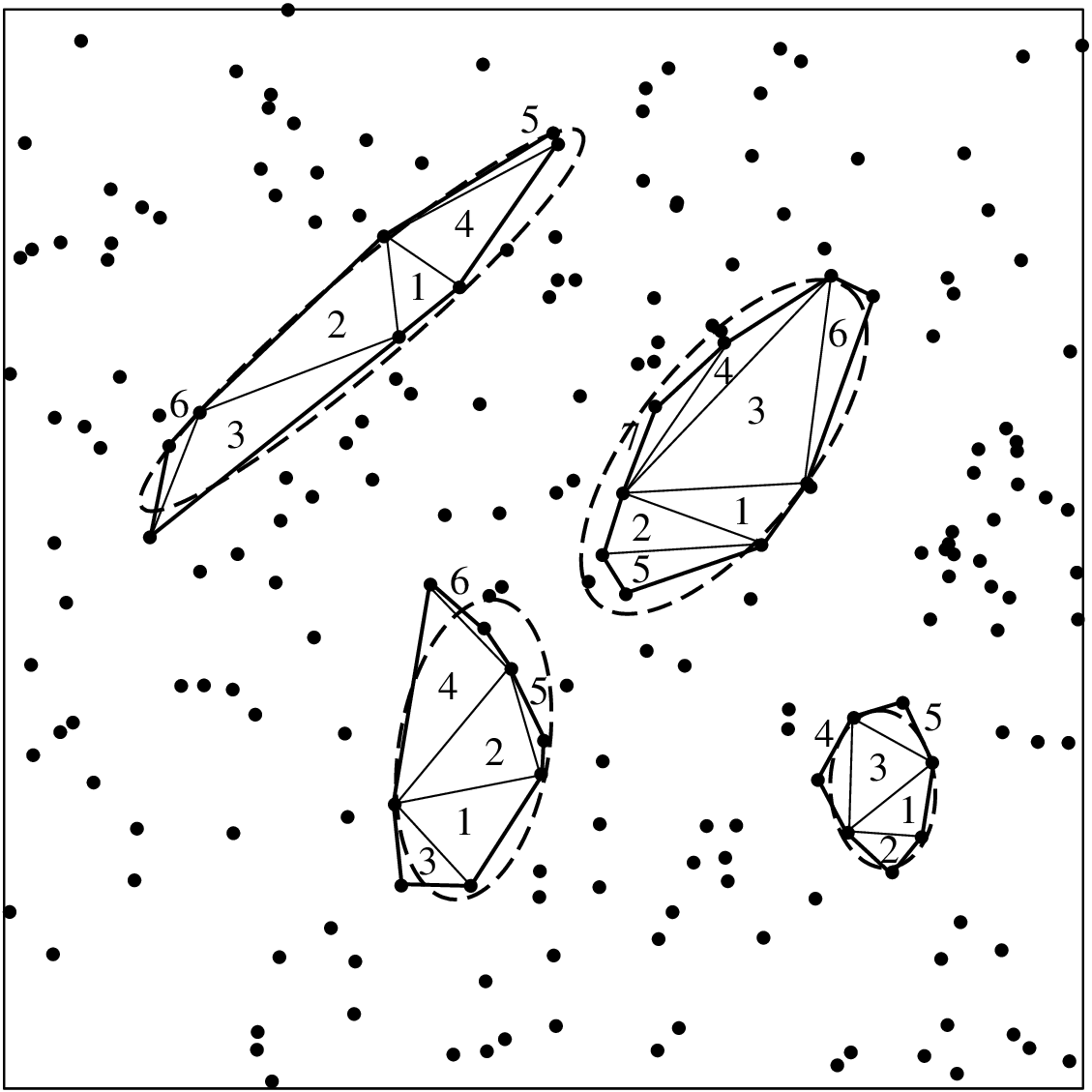}
\vspace{0.6 cm}
\caption{\label{fig7} An illustration of the method of construction of
the elliptic holes in the random configuration of the obstacles (see text
for details).  The numbers indicate the order in which each triangle is
added until the final polygon, defined by the thick segments, is formed.
The corresponding ellipses of inertia are also shown.}
\end{figure}

In order to understand the physical meaning of the observed crossover
lengths, we try to relate them with some characteristic length present
in the host medium.  For this, it must be taken into account that a set
of points distributed randomly in the plane does not show a homogeneous
configuration, as e.g., in a liquid. On the contrary, it consists of
clusters of close points and big holes, see Fig.~\ref{fig7}. It can be expected that the size
and shape of these holes are related to the relaxation dynamics of the
rods at short and intermediate time scales since they will determine how
far the rod can propagate without hitting an obstacle.  Typically such
holes are not circular but rather elliptic, as can be recognized from
Fig.~\ref{fig7}. Due to the strong disorder present in the configuration
of the obstacles, one should expect a wide distribution for the size
of the holes and in the following we present a well defined method to
determine such a distribution. The algorithm goes as follows:
\\\\
1) Pick a random point $Q$ in the plane. (Note that with probability 
one this will not be an obstacle.)\\
2) Look for the three nearest obstacles $O_{1},O_{2},O_{3}$ that enclose $Q$.\\
3) Select (randomly) one of the three sides $O_{i}O_{j}$ of the triangle 
$O_{1}O_{2}O_{3}$.\\
4) Choose randomly one of the remaining obstacles, $O_{\rm new}$, with the conditions that
i) The triangle $O_{i}O_{j}O_{\rm new}$ does not enclose any other obstacle and
ii) The polygon $O_{1}O_{2}O_{3}O_{\rm new}$ has only internal angles $\psi < \pi$.\\
5) If i) and ii) are fulfilled, the initial triangle is expanded to
the new polygon $O_{1}O_{2}O_{3}O_{4}$, with $O_{4}\equiv O_{\rm new}$.  If not,
$O_{\rm new}$ is removed as a candidate to expand the initial triangle from
the side $O_{i}O_{j}$, and we repeat step 3) and 4) until we are
able to expand the initial triangle.\\ 
We now repeat steps $3)\rightarrow 5)$ (note that now in step 3) we
have a polygon of order $n > 3$ instead of a triangle) until i) and ii)
are not fulfilled by anyone of the remaining obstacles and the polygon
$O_{1}O_{2}...O_{n}$ cannot be expanded any more. Thus this procedure
has generated the ``largest'' possible convex polygon that has $Q$ in its
interior and obstacles at its corners. Note that each point $Q$ will have
several such polygons, since steps 3) and 4) contain a random element.

To characterize the shape of each of these closed polygons we calculate its
tensor of inertia (by taking mass unity for each corner of the polygon).
The two eigenvalues of this tensor can be used to define the
ellipse of inertia, which in turn can be used (by means of $A_{\rm max}/2$
and $A_{\rm min}/2$, its major and minor semi-axis, respectively) to
characterize the shape of the elliptic hole associated with the polygon.

An illustration of this procedure is given in Fig.~\ref{fig7}.
As mentioned above, the resulting polygon (and associated ellipse of inertia)
to a given point $Q$ are generally not unique, and different
random selections of the lower-order polygon sides and of the remaining
obstacles in the consecutive upgrading steps can lead to a different final
polygon. This arbitrariness finds its corresponding part in the dynamics
of the rods in that there the trajectory depends not only on the initial
location of a rod but also on the direction of its initial velocity.
However it must be stressed that we are investigating quantities as
diffusion constants or relaxation times that are obtained as averages
over an {\it ensemble} of rods. Thus we will try below to relate the
observed features for these latter quantities with the {\it average} size
of the elliptic holes in the host medium. 

\begin{figure}
\centering
\vspace{1 cm}
\includegraphics[width=0.45\textwidth]{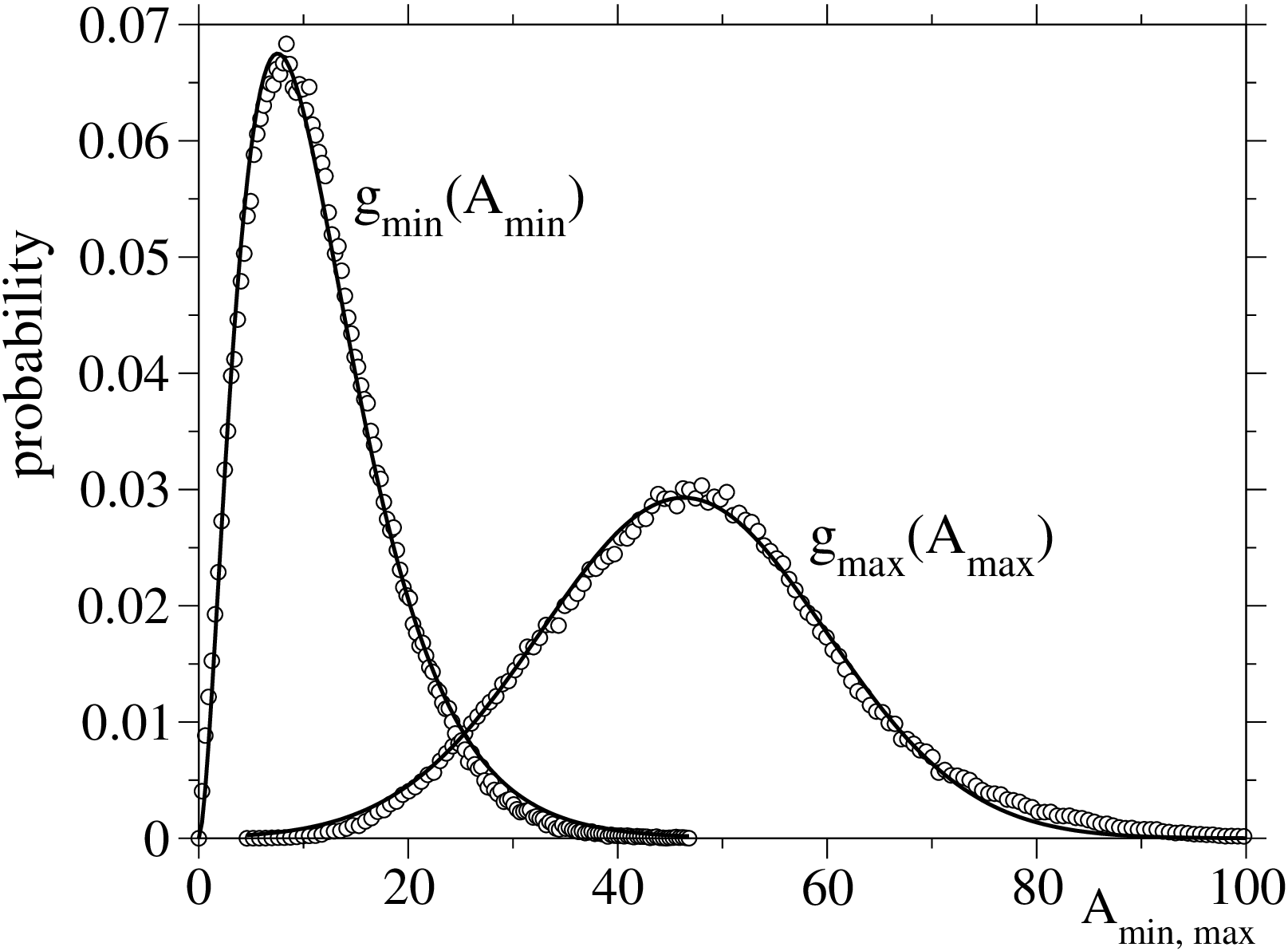}
\caption{\label{fig8} Points: Distribution of the minor and major axis
of the elliptic holes for a density $\rho = 10^{-2}$. Lines are fits
to a Gamma and to a Gaussian function, respectively.}
\end{figure}

To measure the size and shape of the local elliptic holes we used,
for a large number of random points $Q$, and a few times for each
$Q$, the above described algorithm. 
This allowed us to obtain the distribution of the minor, $g_{\rm
min}(A_{\rm min})$, and major axis, $g_{\rm max}(A_{\rm max})$,
which are shown in Fig.~\ref{fig8} for a density $\rho= 10^{-2}$.
The distribution $g_{\rm min}(A_{\rm min})$ is rather asymmetric and
shows a noticeable tail. It can be well described (see Fig.~\ref{fig8})
by a Gamma distribution $g_{\rm min}(A_{\rm min})=(z/e)^{z}(A_{\rm
min}/\bar{A}_{\rm min})^{z-1} \exp[-z(A_{\rm min}-\bar{A}_{\rm
min})/\bar{A}_{\rm min}](\Gamma(z)\bar{A}_{\rm min})^{-1}$.  In this
expression $\bar{A}_{\rm min}$ is the average minor axis, $z$ is a
dimensionless shape parameter, and $\Gamma(z)$ is the Gamma
function for $z$~\cite{abramowitz64}. The fitting procedure yields the
values $\bar{A}_{\rm min}=12$ and $z=2.7$. The standard deviation,
given by $\sigma=\bar{A}_{\rm min}/\sqrt{z}$, is $\sigma=7$. On the
other hand, the distribution of major axis is rather symmetric and can
be well described by a Gaussian $g_{\rm max}(A_{\rm max})=\exp[-(A_{\rm
max}-\bar{A}_{\rm max})^2/2\sigma^{2}]/\sqrt{2\pi}\sigma$, with an
average value $\bar{A}_{\rm max}=47$ and a standard deviation $\sigma=14$.
Obviously, for any other density $\rho$ the corresponding average axis
-and analogously for the standard deviations- will be simply obtained
by a scaling relation

\begin{equation}
\bar{A}(\rho)=\bar{A_{0}}(\rho_{0}/\rho)^{1/2}
\label{eq9}
\end{equation}

\noindent
where $A_{0}$ corresponds to the -minor or major- average axis for the
reference density $\rho_{0}=10^{-2}$.  Thus from Eq.~(\ref{eq9})
we find the relations $A_{\rm min}=1.2\rho^{-1/2}$  and $A_{\rm
max}=4.7\rho^{-1/2}$.

We have shown that there are two characteristic length scales in a
random host medium: the averages of the minor and of the major axes of
the elliptic holes. These latter values nicely match 
respectively the first, $\ell_{1}$, and second crossover length, $\ell_{2}$,
observed for the diffusion constants. From these correspondences we
obtain immediately a simple physical picture for the dynamics of the rod.
The system has one rotational degree of freedom and two translational
ones. The latters can be decomposed into components parallel to the
minor and major axes of the local elliptic hole.  For rods of length $L <
\bar{A}_{\rm min}$, rotations and translations are, on average, hindered
only weakly by infrequent collisions with the obstacles. Therefore,
the trajectories will look like those of a typical Brownian particle,
as has been shown in Fig.~\ref{fig1} for $L=3$, and the translational
and rotational dynamics will be basically isotropic.

When crossing the first characteristic length $\bar{A}_{\rm min}$,
the rod, on average, will not be able to pass transversally between
obstacles that have a distance less than $\bar{A}_{\rm min}$, and many
of the relaxation channels that correspond to a motion perpendicular
to the minor axis will be suppressed due to collisions. At the same
time also large angle rotations will be strongly suppressed. These
effects will result in the observed dynamic crossover seen in $D_{\rm
CM}$ and $D_{\rm R}$.

As can be recognized in the trajectory for $L=39$ in Fig.~\ref{fig1}, no big
blobs are present, in contrast to that for $L=3$. Certainly, due to the
hindrance on the transversal motions, rods of the former size
cannot perform narrow closed loops, and big blobs are not formed in the time
scale of the simulation.
The breakdown of the isotropy of the dynamics
also explains why the Stokes-Einstein-Debye relation is violated for
sufficiently long rods.  For supercooled liquids close to the glass
transition temperature, such a violation is explained as the result of
the formation of fluidized domain bottlenecks which are efficient for
the translational diffusion of the molecules but not for rotational
relaxation \cite{stillinger}. A similar picture can be established
for the generalized Lorentz gas: though transversal motion is strongly
hindered, the rod can escape from the local cage formed by the neighbors
by a longitudinal motion, -i.e., by moving through a bottleneck between
neighboring holes. Though this mechanism will result in a significant
spatial decorrelation, it will not change very much the orientation
of the rod, and a complete angular decorrelation will take place only
for a long displacement from the initial position. As a consequence,
translational dynamics will be strongly enhanced in comparison with
rotations, leading to the breakdown of the Stokes-Einstein-Debye relation.

As there is only one rotational degree of freedom, no further crossovers
will be observed for $D_{\rm R}$ before the expected glass transition
at large $L^{2}\rho$.  However, a second crossover will be observed
for $D_{\rm CM}$ when crossing $\bar{A}_{\rm max}$, due to suppression
of relaxation channels through the major axis.  Since now rods are on
average longer than the holes, transversal motion will be strongly
hindered, leading to the quasi-linear paths observed for $L=79$ in
Fig.~\ref{fig1}. It is worthwhile to remark that within this physical
interpretation, in the general $n$-dimensional case where holes are
hyperellipsoids, $n$ dynamic crossovers should be observed for $D_{\rm
CM}$ -though it would probably be difficult to separate them except for
very small densities. In contrast to this the $n-1$ rotational degrees
of freedom (around the $n-1$ principal axis perpendicular to the rod)
are equivalent by symmetry and hence only a single crossover should
be observed for $D_{\rm R}$ also in the general $n$-dimensional case.
Experimental and computational tests of these predictions would certainly
be interesting.

Finally, as we have mentioned above, for sufficiently large values of
$L^{2}\rho$, the system should certainly show a ``glass transition'',
-in the sense of a transition from an ergodic to a non-ergodic state of
zero diffusivity- both for the translational and for the orientational
degrees of freedom, as it is observed for the formers in the original
Lorentz gas.  It is not clear whether the observed breakdown of the
scaling for the translational diffusion constant at the largest values of
$L^{2}\rho$ corresponds to the beginning of a decay to zero through some
MCT-like power law $D_{\rm CM} \sim (\zeta_{\rm c}-L^{2}\rho)^{\chi}$~\cite{gotzea}.
On the other hand, for the rotational diffusion constant no significant
breakdown of the scaling has been observed, suggesting that for rotations
the $L^{2}\rho$-investigated range is still far from the glass transition.
Simulations at larger values of $L^{2}\rho$ should shed light about
this questions. Work in this direction is in progress. 

It must be remembered that the results presented in this paper
correspond to a random configuration of disk-like obstacles that {\it
can overlap}. If we define a characteristic distance $d=\rho^{-1/2}$,
we find that $d$ is noticeably larger than the soft-disk diameter:
$d > 5\sigma$. Therefore, for the investigated densities, we do
not expect significantly different
results if the configuration of the host medium was slightly modified
such that the obstacles do no longer overlap (e.g. by displacing them by a
small amount). (Differences must be expected at significantly higher densities, since
for the Lorentz gas of hard disks, $N=1$, the details
of the occuring glass transition do indeed depend on the nature of
disorder~\cite{leutheusser}.)  However, if the nature of the host medium
is modified in such a way that the location of the obstacles corresponds
to a liquid-like configuration, i.e.  it shows a new intrinsic length
scale, it might be expected that the resulting dynamics differs from the
one discussed in the present paper. Work in this direction is in
progress.

\section{Conclusions}

We have investigated, by means of molecular dynamics simulations, the
dynamics of a rod in a random frozen environment for a wide range of rod
length and density of obstacles.  The calculated diffusion constants for
rotation and translation of the center of mass can be scaled onto a master curve
that has been deduced from simple scaling and kinematic arguments. The
abrupt crossovers observed in the dynamic exponents take place when the
rod length crosses two characteristic length scales: the average minor
and major axis of the elliptic holes present in the random host medium.
In analogy to the observation for supercooled liquids close to the glass transition,
the Stokes-Einstein-Debye relation is violated for rods longer than the
average minor axis of the elliptic holes. It is found that this effect can be
understood by the longitudinal
motion of the rod that leads to a fast decorrelation of the translational degrees of freedom 
but is inefficient for the rotational one.

\acknowledgments
We thank E. Frey for useful discussions.  A.J.M. acknowledges financial
support from the Basque Government. Part of this work was supported by
the European Community's Human Potential Program under contract
HPRN-CT-2002-00307, DYGLAGEMEM.

%
%
%

%

%

%
%
\vspace{1 cm}
\vspace{1 cm}
\vspace{1 cm}
\vspace{0.7 cm}
%
%
%
%
%
%

\vspace{-0.2 cm}
\begin{thebibliography}{99}
\vspace{-1 cm}
\bibitem{lorentz} H. A. Lorentz, Arch. Neerl. {\bf 10}, 336 (1905).
\bibitem{bruin} C. Bruin, Phys. Rev. Lett. {\bf 29}, 1670 (1972).
\bibitem{alder} B. J. Alder and W. E. Alley, J. Stat. Phys. {\bf 19}, 341 (1978).
\bibitem{gotzea} W. G\"{o}tze, E. Leutheusser, and S. Yip, Phys. Rev. A {\bf 23}, 2634 (1981).
\bibitem{gotzeb} W. G\"{o}tze, E. Leutheusser, and S. Yip, Phys. Rev. A {\bf 24}, 1008 (1981).
\bibitem{gotzec} W. G\"{o}tze, E. Leutheusser, and S. Yip, Phys. Rev. A {\bf 25}, 533 (1982).
\bibitem{masters} A. Masters, and T. Keyes, Phys. Rev. A {\bf 26}, 2129 (1982).
\bibitem{keyes} T. Keyes, Phys. Rev. A {\bf 28}, 2584 (1983).
\bibitem{machta} J. Machta and S. N. Moore, Phys. Rev. A {\bf 32}, 3164 (1985).
\bibitem{binder} P. M. Binder and D. Frenkel, Phys. Rev. A {\bf 42}, R2463 (1990).
\bibitem{cicerone95a} M. T. Cicerone, F. R. Blackburn, and M. D. Ediger,
J. Chem. Phys. {\bf 102}, 471 (1995).
\bibitem{cicerone95b} M. T. Cicerone and M. D. Ediger, J. Chem. Phys. {\bf 103}, 5684 (1995).
\bibitem{andreozzi97} L. Andreozzi, A. di Schino, M. Giordano, and D. Leporini,
Europhys. Lett. {\bf 38}, 669 (1997).
\bibitem{taschin01} A. Taschin, R. Torre, M. Ricci, M. Sampoli, C. Dreyfus, and R. M. Pick,
Europhys. Lett. {\bf 56}, 407 (2001).
\bibitem{moreno} A. J. Moreno and W. Kob, Philos. Mag. (in press); cond-mat/0303510
\bibitem{doia} M. Doi, J. Phys. (Paris) {\bf 36}, 607 (1975).
\bibitem{doib} M. Doi and S. F. Edwards, J. Chem. Soc. Faraday Trans. {\bf 74}, 560 (1978).
\bibitem{fixman} M. Fixmann, Phys. Rev. Lett. {\bf 55}, 2429 (1985).
\bibitem{teraoka} I. Teraoka, N. Ookubo, and R. Hayakawa, Phys. Rev. Lett. {\bf 55}, 2712 (1985).
\bibitem{szamel} G. Szamel, Phys. Rev. Lett. {\bf 70}, 3744 (1993).
\bibitem{egelstaff} P. A. Egelstaff, {\it An Introduction to the Liquid State}
(Oxford University Press, Oxford, U.K., 1994).
\bibitem{fujara} F. Fujara, B. Geil, H. Sillescu, and G. Fleischer, Z. Phys. B {\bf 88}, 195 (1992).
\bibitem{ediger} M. D. Ediger, Ann. Rev. Phys. Chem. {\bf 51}, 99 (2000).
\bibitem{abramowitz64}
M. Abramowitz and I. A. Stegun, {\it Handbook of Mathematical Functions} (Dover, New York, 1964).
\bibitem{stillinger} F. H. Stillinger and J. A. Hodgdon, Phys. Rev. E {\bf 50}, 2064 (1994).
\bibitem{leutheusser} E. Leutheusser, Phys. Rev. A {\bf 28}, 2510 (1983).
\end{thebibliography}
\end{document}